\documentclass[12pt]{article}
\usepackage{graphicx,amsmath,amssymb}
\def\P{I\!\!P}

\newlength{\dinwidth}
\newlength{\dinmargin}
\def\lapproxeq{\lower .7ex\hbox{$\;\stackrel{\textstyle                                                    
<}{\sim}\;$}}                                                    
\def\gapproxeq{\lower .7ex\hbox{$\;\stackrel{\textstyle                                                    
>}{\sim}\;$}}                                                    
\def\be{\begin{equation}}                                                    
\def\ee{\end{equation}}                                                    
\def\bea{\begin{eqnarray}}                                                    
\def\eea{\end{eqnarray}}

\begin{document}
\begin{flushright}                                                    
IPPP/19/40  \\                                                    
\today \\                                                    
\end{flushright} 
\vspace*{0.5cm}
\begin{center}
{\Large\bf The fusion of hard and soft Pomerons:  }\\
\vspace{.5cm}
{\Large\bf 3-jet diffractive production}\\

\vspace{1cm}
V.A. Khoze$^{(a,b)}$, A.D. Martin$^{(a)}$, M.G. Ryskin$^{(a,b)}$, A.G. Shuvaev$^{(b)}$ and 
I.V. Surnin$^{(b,c)}$ \\

\vspace{.5cm}
$^{(a)}$ Institute for Particle Physics Phenomenology, University of Durham, Durham, DH1 3LE \\                                                   
$^{(b)}$Petersburg Nuclear Physics Institute, Kurchatov National
Research Centre,
Gatchina, St. Petersburg 188300, Russia\\
$^{(c)}$St. Petersburg State University, St. Petersburg, Russia

\end{center}

\vspace{0.5cm}
\begin{abstract}
\noindent
We consider the central exclusive production
of high $E_T$ jets $pp\to p+(X+{\rm dijet})+p$. In particular we study the possible contamination of the purely exclusive signal by semi-exclusive production where no
other secondaries are emitted in one hemisphere, between the highest
$E_T$ dijet and the recoil proton, while in the other hemisphere a
third jet, plus possibly additional hadron activity, is allowed, but
still separated
from the incoming proton by a large rapidity gap.
This process arising from the fusion of a hard and a soft Pomeron has not been considered before. It turns out that it gives a negligible contribution. 
The calculation involves a careful treatment of the QCD colour structure of the amplitudes. 
\end{abstract}

\section{Introduction}
Central Exclusive Production (CEP) of high $E_T$ jets is of interest for at
least two reasons. First, due to its relatively large cross
section it plays the role of a `standard candle' for the
calculations of different CEP cross sections~\cite{St-Cand};
in particular, for the evaluation of the chance to observe new Beyond the Standard Model (BSM)
physics in the clean environment provided by CEP kinematics. Next,
due to the $J_z=0$ selection rule~\cite{J0}, the production of 
quark jets is strongly suppressed by a factor $m^2_q/E^2_T$
(where $m_q$ is the quark mass). Therefore in the CEP process we observe the gluon jets with a good
purity\footnote{ In principle, away from the exact forward region
there is a $|J_z|=2$ contribution, see e.g. \cite{J0}.
However an explicit calculation in \cite {dur-rev} shows that the contribution of 
 such a term is very small $\sim 1$\%.}.
 Thus we have a `gluon factory'
which provides an excellent possibility to study gluon
jets \cite{Pros}.

Exclusive dijet production $pp\to p+{\rm dijet}+p$ is
shown symbolically  in Fig.1a.
Within the perturbative QCD approach the Pomeron may be described
at lowest order in $\alpha_s$ by the two gluon exchange diagram,
and we are led to diagram Fig.1b.  At the lowest $\alpha_s$ order
the CEP dijet cross section was calculated in~\cite{Col}.
More precise results accounting for the leading logarithmic
corrections  were obtained in~\cite{KMR-dijet}
(see also the reviews~\cite{dur-rev,dur-rev1}).

However experimentally it is challenging to observe only two jets,
without any additional secondaries. As a rule besides the two high
$E_T$ jets there are other particles, with smaller transverse
momenta, $p_t$, and it is not quite clear whether these particles
were produced during the jet hadronization or whether they must be
considered as an additional relatively low $E_T$ jet. Moreover, it
is not excluded that some low $p_t$ particles were missed by the
detector.

\begin{figure}
\vspace{-1cm} 
\hspace{0.8cm}
\includegraphics[scale=0.32]{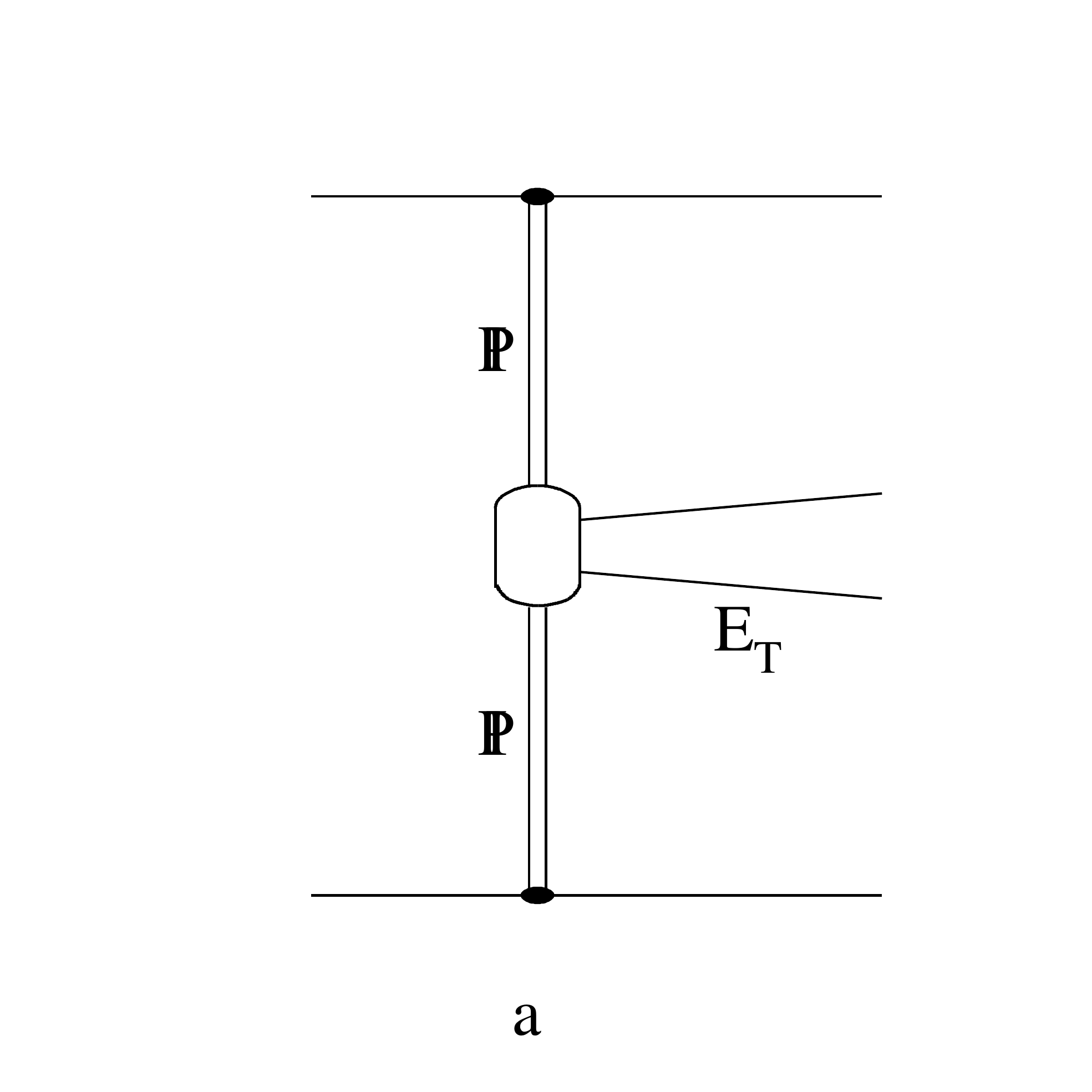}
\hspace{-2.3cm}
\includegraphics[scale=0.32]{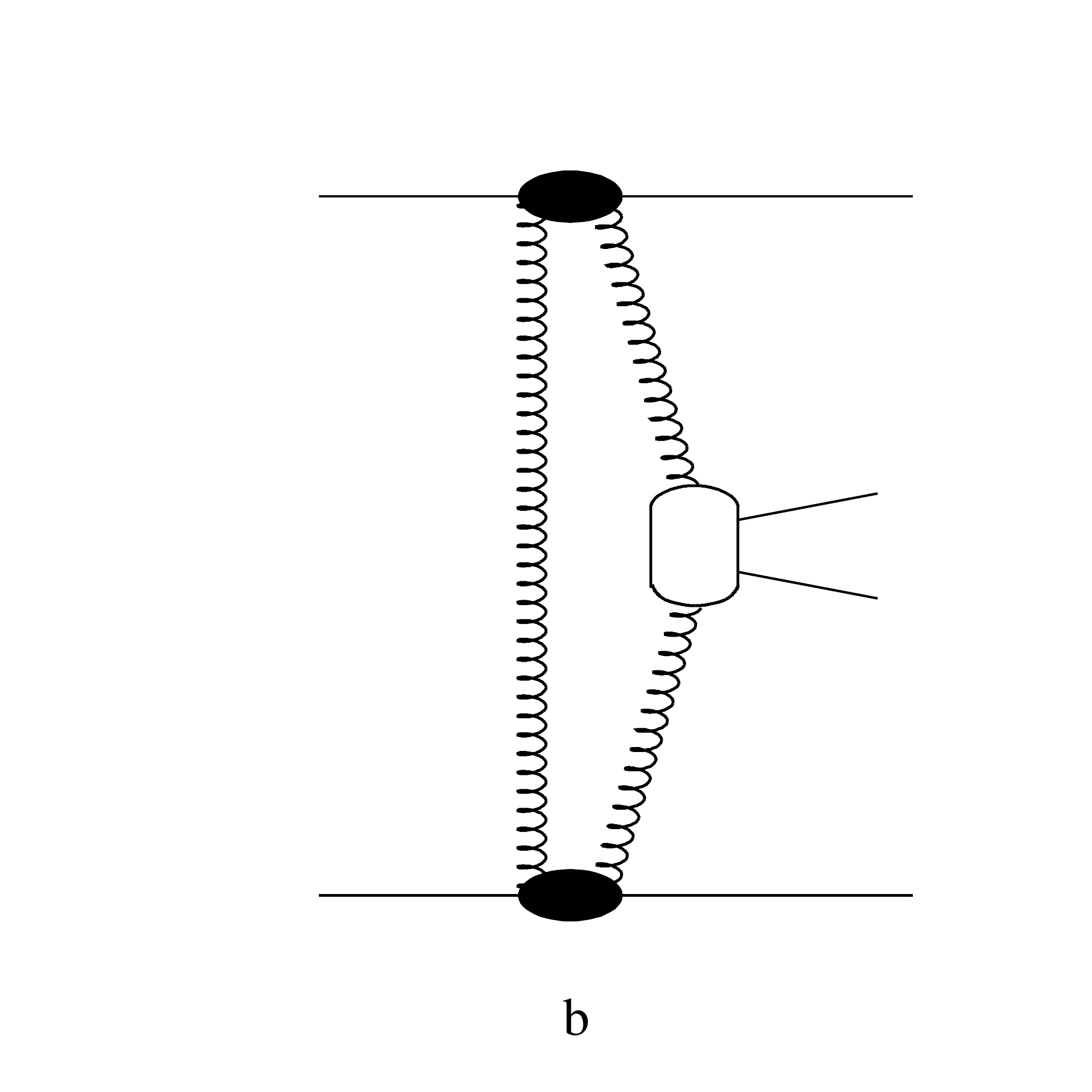}
\hspace{-2.3cm}
\includegraphics[scale=0.32]{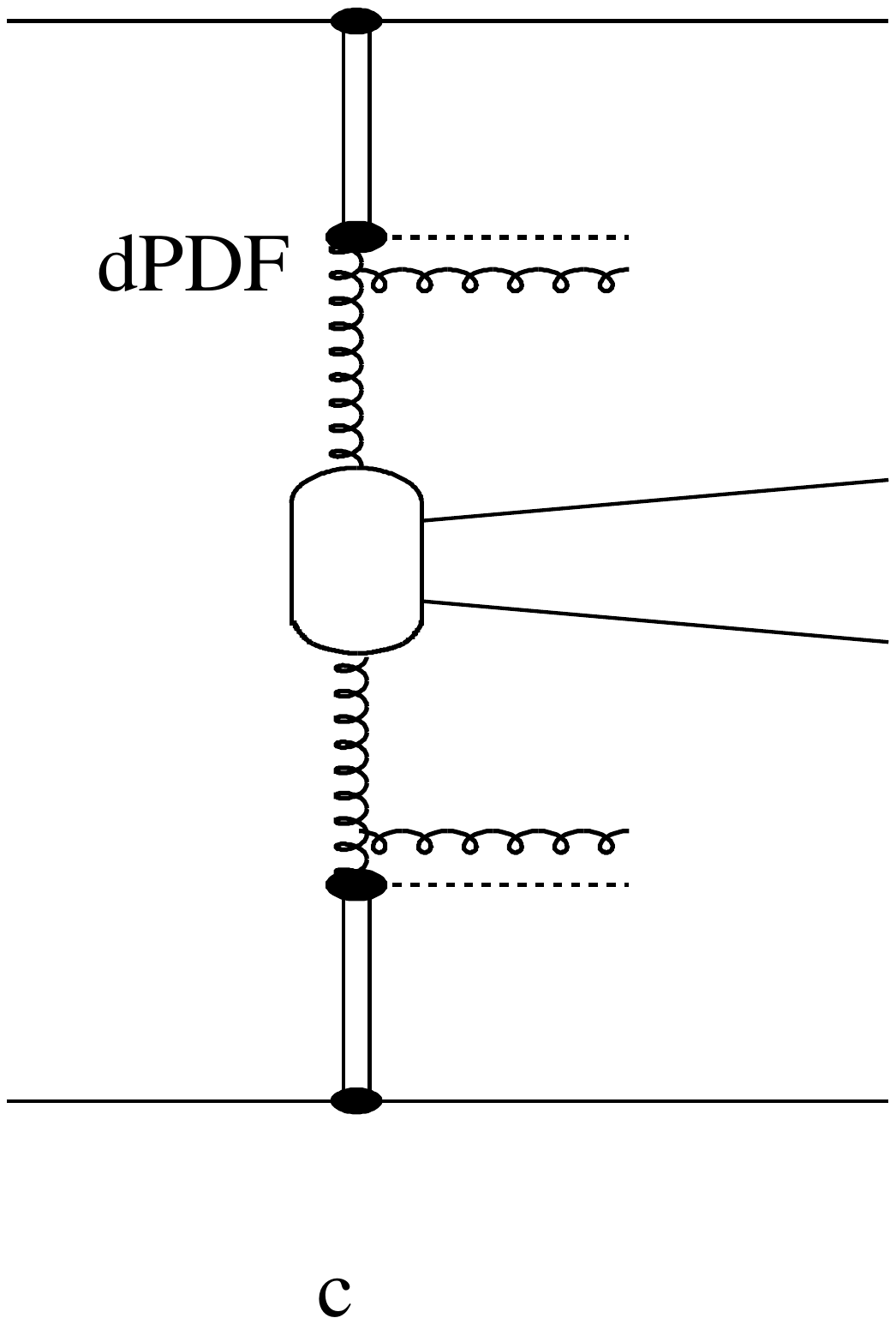}
\vspace{-.3cm} \caption{\sf Exclusive (a,b) and semi-exclusive (c)
high $E_T$ dijet production via Pomeron-Pomeron fusion; in the
case of diagram (b) the perturbative QCD Pomeron is represented schematically  by the
two gluon exchange.}
\end{figure}
For example, in order to  select pure CEP dijets in the CDF
experiment~\cite{CDF} the ratio , $R_{jj}$, of the dijet mass,
$M_{jj}$, to the mass of the whole central system, $M_X$ was
plotted. Pure CEP events should correspond to a peak at
$R_{jj}=M_{jj}/M_X=1$. Unfortunately the peak at $R_{jj}=1$ is not well manifested
 on the top of the background caused by Double
Pomeron Exchange (DPE) contribution\footnote{See the discussion in \cite{DPE}. }, that is inclusive dijet
production in Pomeron-Pomeron collisions  (sketched in Fig.1c). The peak looks more like a shoulder in $R_{jj}$ distributions of events with $E_T>10$ GeV (Fig.14a of \cite{CDF}) and only one point which exceeds the background by 1.5$\sigma$ can be seen in Fig.14c for $E_T>25$ GeV.

 DPE production can be described in terms of diffractive parton
 distributions (dPDF). The dPDF, that is the distribution of partons
 inside the Pomeron, was measured at HERA by selecting Deep Inelastic Scattering (DIS) events with
 a Large Rapidity Gap (LRG) between the incoming proton and the hadron
 system produced by a heavy photon (or by selecting events with a leading proton
 carrying away a large momentum fraction $x_L\to 1$)~\cite{H1,Zeus}.
 The  cross  section of DPE dijet production is given by the convolution
 of the `hard' $2\to 2$ matrix element squared with the parton distributions
 originating from the two Pomerons. It was calculated in~\cite{Col,PR}
 and implemented in Monte Carlo generators, like e.g.  POMWIG~\cite{POMW}.

 As seen from Fig.1c, while pure CEP events at the parton level
 have only two high $E_T$ jets, for a DPE process at least
 four partons/jets are produced - the two high $E_T$ jets together with two spectators
 which are needed to compensate the colour of the parton extracted from the incoming
 Pomeron; besides this there may be other partons (shown by the dotted lines
 in Fig.1c) radiated during the dPDF evolution. For this reason the ratio
 $R_{jj}<1$ for DPE events.

 The aim of this paper is to  consider an `intermediate' configuration between the CEP and DPE
 possibilities -- that is dijet production due to the collison of a parton from the dPDF
 of a `soft' Pomeron on one side with the CEP-like `hard' Pomeron on the other side.
 At the lowest $\alpha_s$ order this corresponds to three-jet production --
 a pair of high $E_T$ jets and a jet-spectator from the soft Pomeron, see Fig.2a.
\begin{figure}
\vspace{-5cm} 
\hspace{-.5cm}
\includegraphics[scale=0.5]{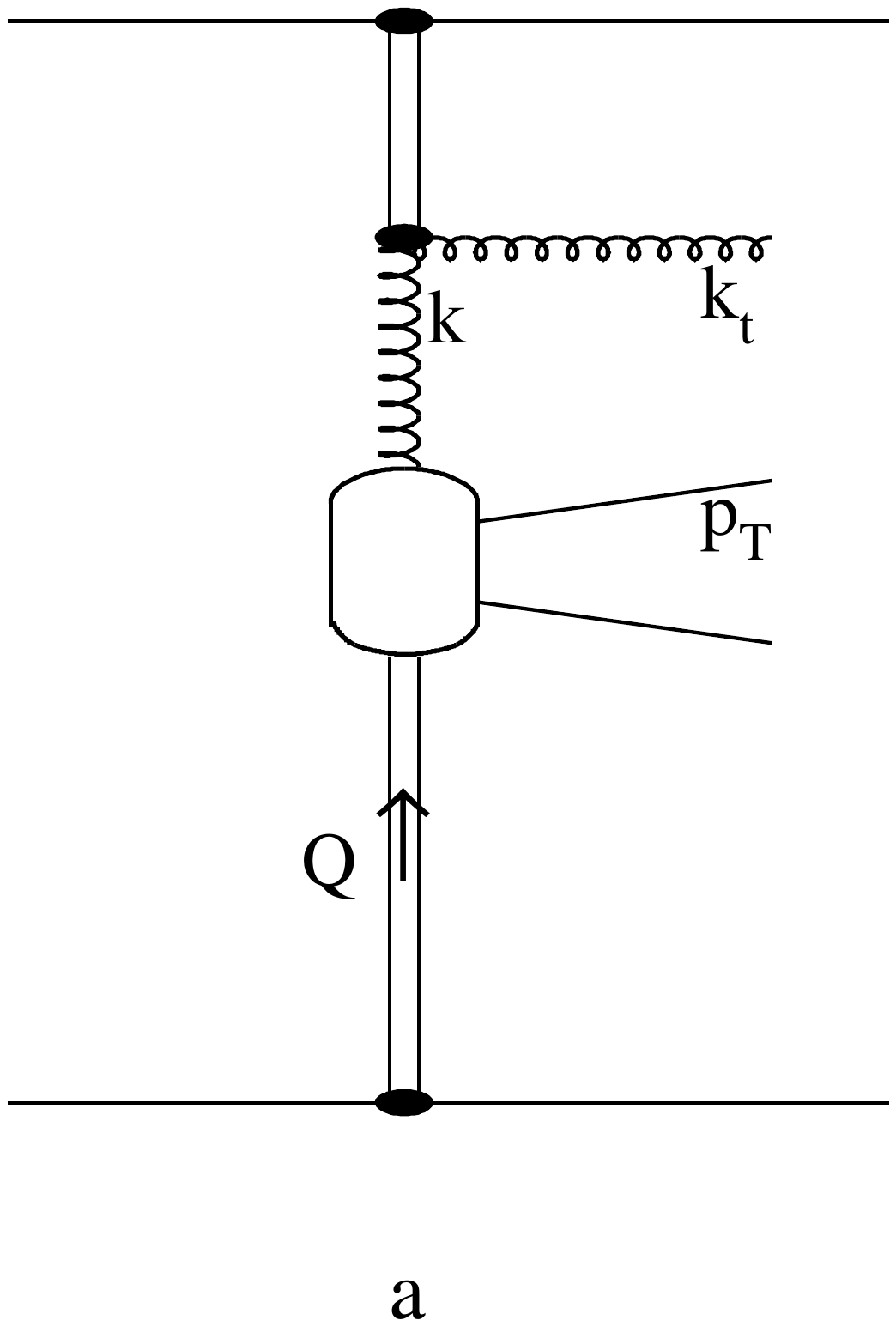}
\hspace{-2.3cm}
\includegraphics[scale=0.5]{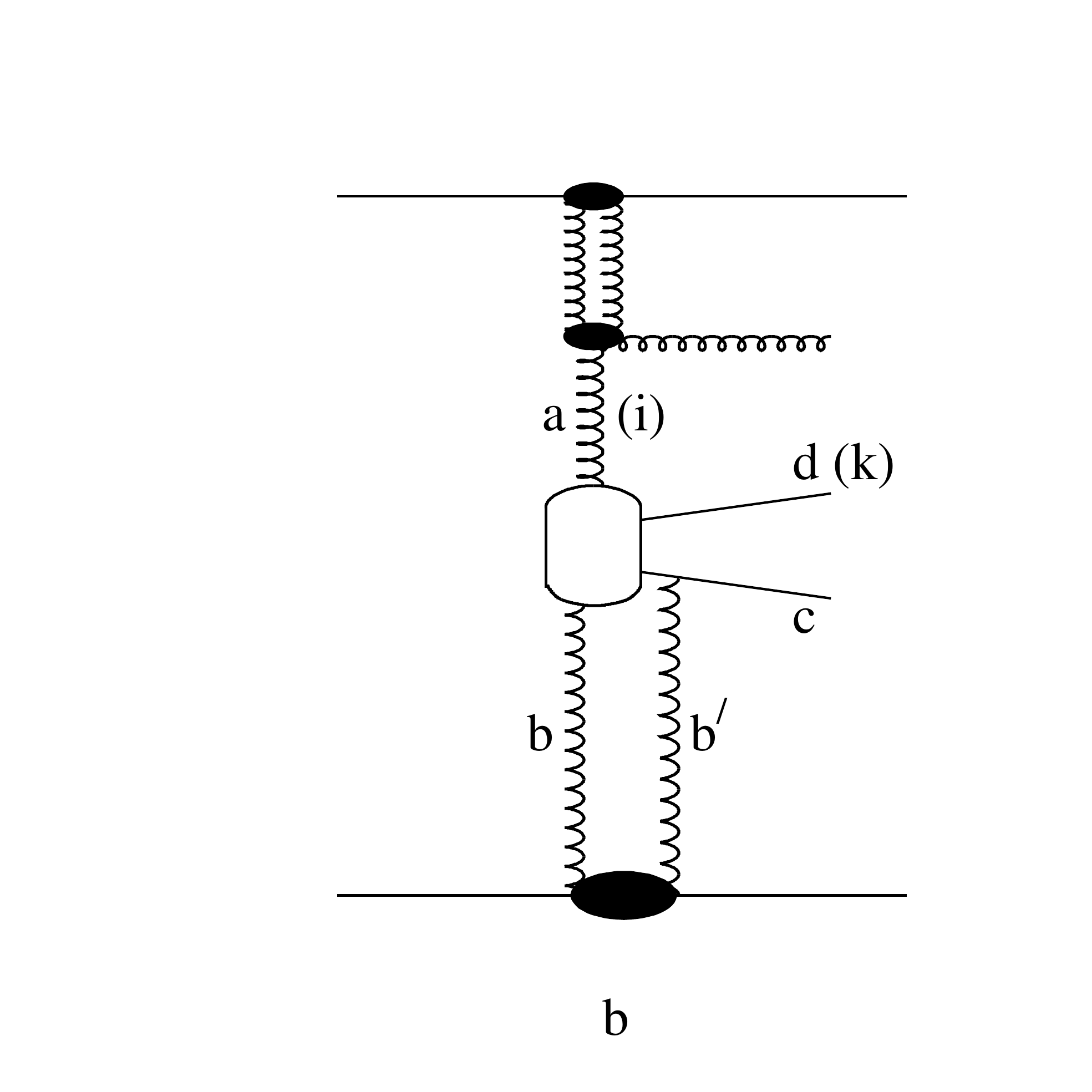}
\vspace{-.3cm} \caption{\sf Diagrams for three jet production via
the fusion of a 'soft' (upper) and the 'hard' Pomeron. Two highest
$p_T$ jets are produced in the fusion of a 'hard' Pomeron with the
 parton (shown as a gluon) from the soft Pomeron PDF. The momenta transferred to the hard matrix element by the hard Pomeron ($Q$) and the parton from the soft Pomeron ($k$) together with the momenta of produced jets are indicated in Fig.2a while the colour indices of the incoming and outgoing gluons (quarks) are shown in Fig.2b.}
\end{figure}

 Strictly speaking an analogous three-jet configuration can be produced
 in pure CEP events as well. One has just to consider the $gg\to 3$-jet
 hard subprocess. For a pure CEP case this was done in~\cite{HLKR-3j}. Another approach was used in~\cite{Lonbl}. In this paper the configuration where the third jet has a relatively small momentum jet ($k_t \ll p_T$) was considered. However here it was not the pure 
 exclusive kinematics since the cross section includes the processes in which additional (softer) jets with $k_t>1.5$ GeV can be radiated.
 
 Note that when the intercept of the Pomeron trajectory, $\alpha_P(0)$,
 is close to 1 there is practically no interference between the pure CEP(3-jet)
 and the CEP(2-jet)$\otimes$DPE amplitudes. Soft Pomeron exchange in the DPE amplitude
 produces an additional (imaginary) factor $i$. Thus we may consider
 the pure CEP(3-jet) and the CEP(2-jet)$\otimes$DPE cross sections separately.

The outline of this paper is  as follows. In section 2 we describe
the detailed structure of the CEP$\otimes$DPE amplitude and show that
it can be `factorized'. That is, the result can be written  as the
convolution of a hard $2\to 2$ matrix element and an integral over
much smaller transverse momenta. 
  Next, in section 3, we obtain an expression for the effective luminosity, corresponding to `hard'-to-`soft' 
  (CEP$\otimes$DPE) Pomeron-Pomeron fusion, for this semi-exclusive process and to describe its main elements, in particular, the Sudakov
 $T$-factor and incoming parton distributions. Recall that this effective luminosity includes all the components which are driven by the relatively soft scale fixed by the transverse momentum, $k\ll E_T$, of a relatively soft third jet. 
We discuss 
 different ways to introduce the infrared cutoff, either by
an effective gluon mass or by the Sudakov $T$-factor and the
incoming unintegrated parton distributions. The hard dijet cross
sections of the relevant subprocesses are given in section 4 while
the numerical estimates of the expected cross section are presented
in section 5. We conclude in section 6.

\section{Soft-hard factorization}
 Here by factorization we mean that we can separate the calculation of
the hard $2\to 2$ subprocess cross section for exclusive production of the high $E_T$ dijet from that
for semi-exclusive production involving low $k_t$ jets. In the calculation of the latter, which includes the incoming parton densities, we introduce an `effective luminosity', see section~\ref{sec:3.3}.

This effective luminosity is formed at a rather low scale driven by the relatively small transverse momentum, $k_t$, of the third jet. On the other hand the hard dijet cross section occurs
at a large ($\sim E_T$) scale. The Sudakov factor, which accounts for the probability not to radiate additional partons in the interval between
$k_t$ and $E_T$, is included in the `effective' luminosity.

The amplitude of CEP$\otimes$DPE dijet production is shown
symbolically in Fig.2a and 2b where both the soft and hard Pomerons are replaced by
two gluon exchange diagrams. Let us denote the transverse
momentum\footnote{The transverse momenta are shown in Fig.3a} of the parton coming from the upper soft Pomeron by $k$,  and the transverse momenta of left and right gluons which
compose the bottom Pomeron by
$q_1,q_2$. If $Q$ is the transverse momentum
transferred to the recoil proton, then $q_{1,2}=q\pm Q/2$.

\begin{figure}
\vspace{-1cm} 
\hspace{-1.cm}
\includegraphics[scale=0.4]{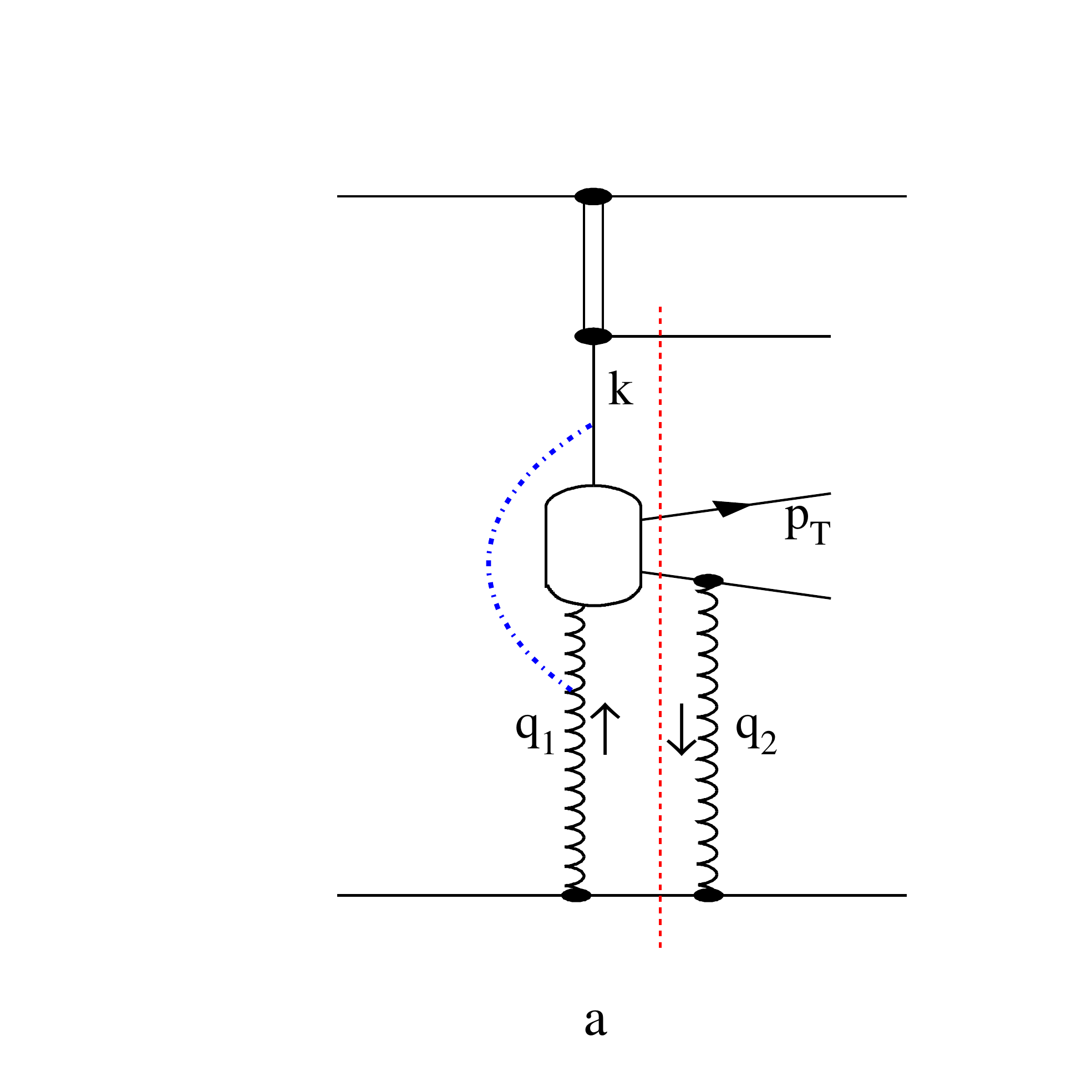}
\hspace{-3.2cm}
\includegraphics[scale=0.4]{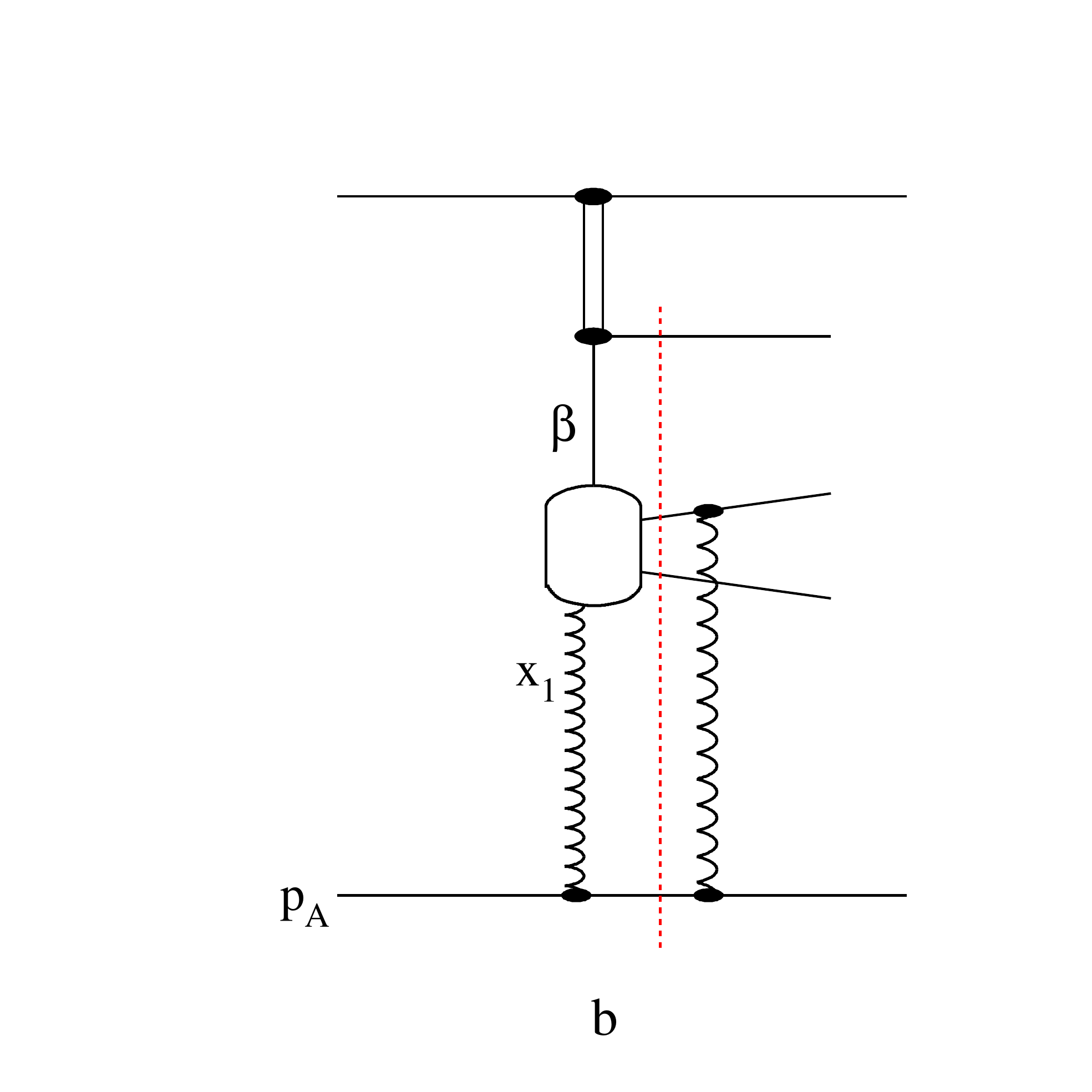}
\hspace{-3.3cm}
\includegraphics[scale=0.4]{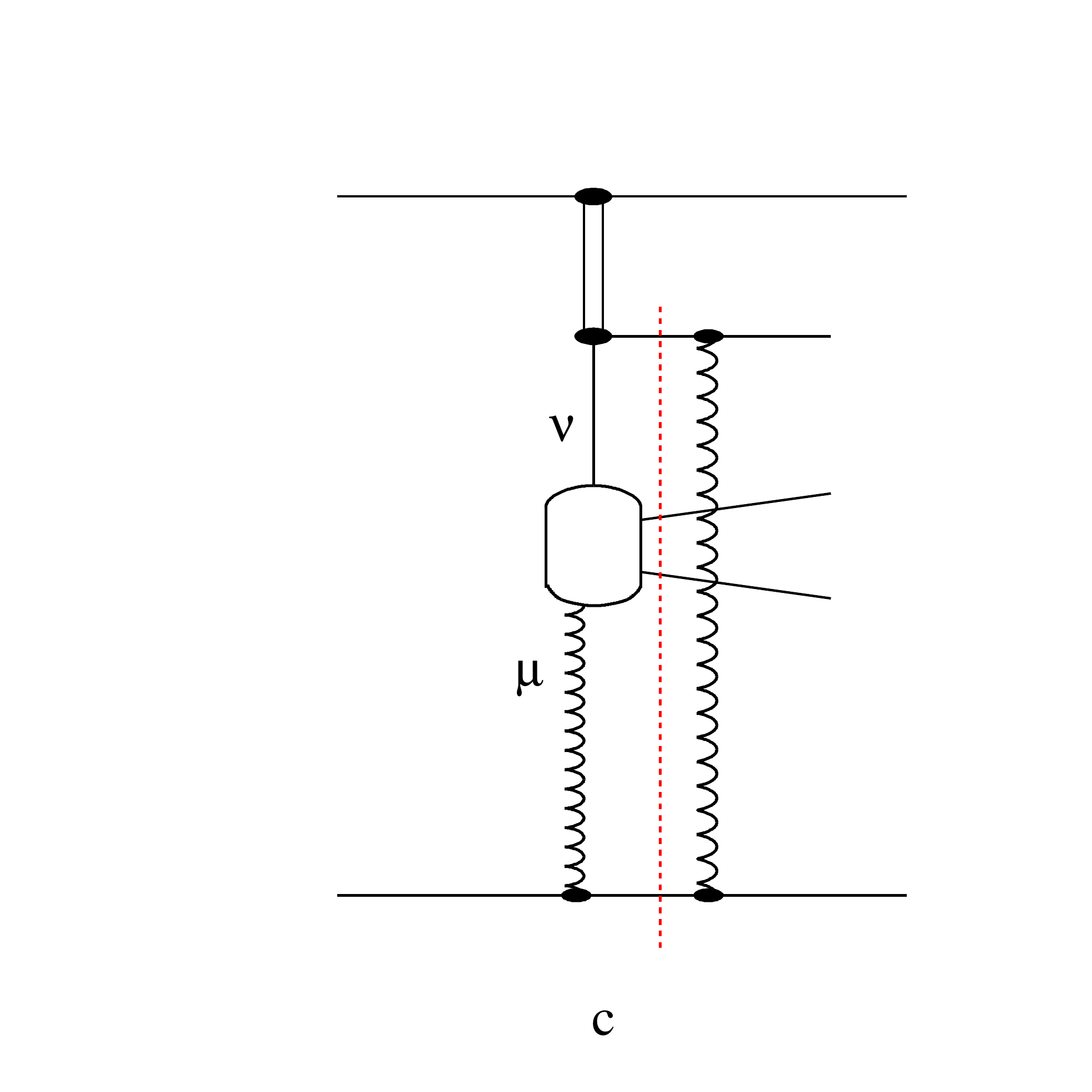}
\vspace{-.3cm} \caption{\sf Three diagrams which describe the
imaginary part of  exclusive 3-jet production amplitude. The momenta, $q_{1,2}$, of incoming gluons from the hard Pomeron and the parton, $k$, from the soft Pomeron are indicated in Fig.3a; $p_T$ is the hard jet momentum. Momentum fractions, $x_1$, and $\beta$ carried by the active gluon in hard Pomeron and by the parton in soft Pomeron are shown in Fig.3b. Lorentz indices of the partons incoming the hard matrix elements are marked in Fig.3c.}
\end{figure}
Since the value of $Q^2$ is limited by the proton form factor the
integrals over $q$ and $k$ (which appear when we calculate the
cross section) are ultraviolet (UV) convergent, while the infrared
(IR) behaviour is regularized by an effective gluon mass,
$m_g=0.6-0.8$ GeV (see e.g.~\cite{PP,CF}) or by the proton size
(in both cases these reflect the confinement effects)  and by the
Sudakov $T$-factor which accounts for the probability not to radiate
additional relatively soft gluons in the  parton-Pomeron fusion process forming
the high $E_T$ dijet. This is the same $T$-factor
which provides the IR cutoff in a pure CEP
amplitude~\cite{KMR-dijet,KMR1}, see section~\ref{sec:2.1}.

Since the essential values of $q$ and $k$ are much smaller than the
high $E_T$ jet transverse momenta, $p_T$, we can neglect $q_{1,2}$
and $k$ when calculating the hard matrix element and use the MHV approach
(see e.g.~\cite{MP}) for the hard $2\to 2$ sub-process amplitude.
Next, recall that in the Pomeron exchange amplitude the imaginary
part dominates, while the real part can be restored (if needed) with
the help of the well known signature factor. That is we 
 consider only the corresponding imaginary part of the amplitude. 

This imaginary
part is given by the sum of three diagrams with the cuts shown in Fig.3 by a vertical dotted
line. That is we have to sum three diagrams: two in which the right gluon couples to
 a high $E_T$ jet and the third in which it couples to the upper Pomeron
parton-spectator. When the right gluon couples to a high $E_T$ jet it
does not affect the kinematics shown in Fig.3a (since $q_2\ll p_T$).
On the other hand in Fig.3c, where the right gluon couples 
to the spectator, we have to replace the parton momentum $k$ by
$k-q_2$.
Note that the final three jet system is colourless. Therefore the
sum of the first two diagrams (Figs.3a,b) has a colour factor equal
(up to the sign)  to that of Fig.3c.\footnote{In the $q_2\ll k$ limit,
when the gluon wavelength $1/q_2$ is much larger than the size of
3-jet system, the gluon $q_2$ probes just the total colour charge of
this colourless system and the interaction amplitude vanishes.}
Thus we can select (factorize) the soft part of the amplitude, $I_{\mu\nu}$  which includes the propagators of two (lower) gluons from the hard Pomeron, the propagator of the parton from soft Pomeron and the corresponding polarization vectors. This is the central factor of the 'effective' luminosity which  should be
 further multiplied by the differential hard dijet cross section as it will be explained in section 3.3. 
\begin{equation}
\label{1} I_{\mu\nu}=\int
d^2q\frac{q_{1\mu}}{(q^2_1+m^2_g)(q^2_2+m^2_g)}
\left[\frac{k_\nu}{k^2+m^2_g}
-\frac{(k-q_2)_\nu}{(k-q_2)^2+m^2_g}\right]T(q,M_{jj})\ .
\end{equation}
We include in (\ref{1}) the Sudakov factor $T$ which accounts for the absence of radiation of an additional jet since it may strongly affect the infrared behaviour of the integral. This factor is given explicitly in the next subsection.
 The transverse indices  $\mu,\nu=1,2$ correspond to incoming gluon polarizations which then should be convoluted with the hard matrix element (shown
  by blob in Fig.3). In the left part of the amplitude we choose the gluon
polarization vectors to be
\be
\vec e_{1,\mu}\simeq -\vec q_{1,\mu}/x_1
~~~~~{\rm and}~~~~~ \vec e_{k,\nu}\simeq -\vec k_\nu/\beta,
\ee
 where $x_1$ is the
lower proton momentum fraction carried by gluon $q_1$ while $\beta$
is the upper Pomeron momentum fraction carried by the parton $k$~\footnote{Factors $1/x_1$ and $1/\beta$ are not included in $I_{\mu\nu}$ but we will account for them in sect.3.3 when calculating the effective luminosity (\ref{Ls}).}
This means that we are working in an axial/planar gauge or using 
Gribov's gauge trick\footnote{Recall that the  left (with respect to
the cut) part of the diagram is gauge invariant.} replacing the (lower) proton 4-momentum, $p_{A,\mu}$, by
$-q_{1,\mu}/x_1$.

For the second (soft in $x$) gluon in the right part of the amplitude of Fig.3 it is better to use Coulomb
polarization, that is to use $g_{\alpha\beta}$ as the spin part of gluon $q_2$ propagator. In this case in the upper vertex of the gluon $q_2$ we have just $p_{A,\alpha}$. This simplifies the calculation of
imaginary part of the
diagram, that is the 
`cut' between the `soft'(right) and the `hard'
(left) gluons; recall that the hard matrix element is to the left
of the `cut'.
Clearly the integral (\ref{1}) has no UV divergency.\footnote{
The IR contribution is smeared by the effective gluon mass $m_g$ and the $T$-factor.}

The values of $I_{\mu\nu}=I_{\mu\nu}(Q,k)$ can be calculated
numerically. In order to perform the convolution with the hard matrix element calculated in terms of helicity amplitudes we consider three possibilities:  $I^{(0)}$,
$I^{(2)}$ and $I^{(q)}$ corresponding to whether the high $E_T$ dijet is produced by gluon-gluon fusion in either the
$J_z=0$ or $J_z=2$ helicity states or by quark-gluon fusion dijet production  $qg\to qg$. We
introduce  the upper index $I^{(s)}$ (with $s=0,2,q$) in order to
consider in turn the
 convolution of the soft part with the different hard matrix elements
 which may describe either the gluon ($gg\to$ dijet)  production
 in $J_z=0$ or $|J_z|=2$ helicity states or $qg\to qg$ production.
If the parton $k$ is a quark then the square bracket in (\ref{1})
should be multiplied by the Dirac matrix $\gamma_\nu$.
Note that for $gg\to gg$ dijet production, $J_z$ is the projection
 of the spin of the dijet system on the longitudinal (beam) axis,
 that is $J_z$ corresponds to the difference of helicities of the incoming gluons with momenta
 $q_1$ and $k$.

Thus we calculate
\begin{equation}
\label{20} I^{(0)}(k)=\frac 14\int d^2Q (I_{xx}+I_{yy})^2\ ,
\end{equation}
\begin{equation}
\label{22} I^{(2)}(k)=\frac 14\int d^2Q
\left((I_{xx}+I_{yy})^2-4I_{xx}I_{yy}+2I_{xy}I_{yx}
+I_{xy}I_{xy}+I_{yx}I_{yx}\right)\ ,
\end{equation}
\begin{equation}
\label{2q} I^{(q)}(k)=\frac 14\int d^2Q
(I_{xx}I_{xx}+I_{yy}I_{yy}+I_{xy}I_{xy}+I_{yx}I_{yx})\ ,
\end{equation}
where we have used the `projectors' $\delta_{\mu\nu}\delta_{\mu'\nu'}$
for the $J_z=0$ state and
$(\delta_{\mu\nu'}\delta_{\nu\mu'}+\delta_{\mu\mu'}
\delta_{\nu\nu'}-\delta_{\mu\nu}\delta_{\mu'\nu'})$
for the $J_z=2$ state. Here the indices $\mu',\nu'$
belong to the complex-conjugated amplitude. For quark-gluon fusion the projector is $\delta_{\mu\mu'}\delta_{\nu\nu'}$.

Indeed, the projector $\delta_{\mu\nu}$ means that we select the $J_z=0$ state and we have to sum over all possible polarizations of the transverse gluon. That is we obtain $(I_{xx}+I_{yy})$. Doing the same with the complex-conjugated amplitude we obtain (\ref{20}). It is not so strightforward with the $J_z=2$ amplitude but multiplying by $\delta_{\mu\nu}$ one can easily check that the product
$(\delta_{\mu\nu'}\delta_{\nu\mu'}+\delta_{\mu\mu'}
\delta_{\nu\nu'}-\delta_{\mu\nu}\delta_{\mu'\nu'})\cdot\delta_{\mu\nu}=0$, that is this projector does not contain $J_z=0$ state while the remaining states of two gluon system have $|J_z|=2$. Finally in the quark case we just sum up all possible combinations of polarizations.

\section{Soft components of production amplitude}
Here we discuss the various components that are required in the calculation of the cross section for semi-exclusive soft production.

\subsection{The Sudakov $T$ factor  \label{sec:2.1}}
The Sudakov form factor correction originates from diagrams
like that show by the dash-dotted line  in Fig.3a. It accounts for
the fact that for exclusive events we do not allow for standard
bremsstrahlung from the colour-charged incoming gluon (or quark).
Within the leading double logarithmic (DL) approximation the Sudakov
$T$-factor reads
\begin{equation}
\label{T}
T(q,\mu)=\exp\left(-\frac{N_c\alpha_s}{4\pi}\ln^2(\mu^2/q^2)\right)\
\end{equation}
for the case when parton $k$ is a gluon. When parton $k$ is a
quark the colour coefficient $N_c=3$ (for the QCD $SU_c(3)$ group)
must be replaced by $(N_c+C_F)/2=13/6$. Accounting for the one-loop
running QCD coupling
$\alpha_s(q^2)=(4\pi/b_0)/\ln(q^2/\Lambda^2_{QCD})$, expression
(\ref{T}) takes the form
\begin{equation}
\label{Tr}
T_g(q,\mu)=\exp\left(\int_{q^2}^{\mu^2}\frac{d\kappa^2}{\kappa^2}
\frac{N_c\alpha_s(\kappa^2)}{\pi}\ln(\kappa/(\kappa+\mu))\right).
\end{equation}
In practice we use more precise expressions for the quark and gluon $T$-factors. These can be
found in~\cite{MRW}.

The upper limit $\mu$ of the integral is taken to be $M_{jj}$, the mass of the high $E_T$
dijet system (see~\cite{TF}), while the lower cutoff, $q^2$, reflects
the cancellation between the coherent radiation from the $q_1$ and $q_2$
gluons for the emission of an extra gluon with wavelength larger than
the size of the colourless $q_1$ and $q_2$ gluon pair.

Recall that strictly speaking the full $T$ factor correction depends on a
particular jet searching algorithm. Depending on the algorithm some
part of bremsstrahlung emission
may be allowed and included in high $E_T$ jet hadronization.

\subsection{Incoming parton distributions}
Note that, due to the cancellation for $k\gg q$ between the first and
the second terms in the square brackets of (\ref{1}), the relevant values
of momentum $k$ is of the order of $q$. The integral is UV
convergent. The dominant contribution comes from the low $q,k$ domain
with an IR cutoff provided by the $T$-factor (\ref{T}) or (for the case
of not too large $E_T$, when the $T$-factor is ineffective) by
 confinement, or by an effective gluon mass $m_g$.

 Note that the calculation of the diagrams of Fig.3 require knowledge of
 unintegrated parton distributions, $f^a$.  For the `soft' (upper) Pomeron these explicitly
 depend on the transverse momentum $k$, that is $f^a=f^a(x_{\P},\beta,k,\mu;t)$ with $a=g,q$,
 where the arguments are defined below.
  These unintegrated distributions can be obtained
 from the (integrated) dPDFs with the help of the KMR/MRW
 prescriptions~\cite{updf,MRW}.
Actually here we use the simplified form
 \begin{equation}
 \label{unint-k}
 f^a(x_{\P},\beta,k,\mu;t)~=~\frac{\partial}{\partial \ln k^2}
 \left[T_a(k,\mu)xa^D(x_{\P},\beta,k^2;t)\right]\ ,
 \end{equation}
 where $a^D$ is the integrated diffractive parton distribution
(dPDF) and $x=x_{\P}\beta$.  For $a^D$ we take the H1 fit B
parametrization which satisfactorily describes the diffractive DIS
data~\cite{H1}. Note that the corresponding dPDF depends on three arguments --
the proton momentum fraction, $x_{\P}$, carried by the
Pomeron, the Pomeron momentum fraction, $\beta$, carried by
the parton, and the scale $\mu$. Besides this there is the
dependence of the Pomeron flux on the momentum transferred, $t$ (see
the parametrization presented in \cite{H1}).

 The situation for the `hard' (bottom) Pomeron is different.
  When the scale is relatively large, say $q_{1,2}>1-2$ GeV, we have to replace
   the two gluon exchange diagram (i.e. the Low-Nussinov two gluon Pomeron~\cite{LN})
   by the Generalized Parton Distribution (GPD) function~\footnote{See e.g.~\cite{GPD} for review.} of proton.
   That is instead of the usual (unskewed) PDF we deal with the GPD
   since the momentum transferred through our Pomeron is not zero. In particular,
   the momentum fraction $x_2\ll x_1$.
 Since we are looking for LRG events, the value of $x_1\ll 1$ is itself small
 and so the Generalized function GPD can be obtained from the known usual PDF
 \cite{Shuv}. In the limit $x_2\ll x_1$ the ratio $R_g=$GPD/PDF is given by
 \begin{equation}
 \label{Rg}
 R_g=\frac{2^{2\lambda+3}}{\sqrt{\pi}}\frac{\Gamma(\lambda+5/2)}{\Gamma(\lambda+4)}\ ,
 \end{equation}
 when the small $x$ behaviour of gluons is described by $xg(x)\propto x^{-\lambda}$
 and where $\Gamma$ is the Gamma function.



\subsection{Effective luminosity   \label{sec:3.3}}
As shown above, the three jet cross section caused by the
fusion of a `soft' and a `hard' Pomeron can be written as the
convolution of the effective luminosity with the cross section of the hard
subprocess. The non-trivial point is that now the third jet (with the
smaller transverse momentum $k$ distribution) is included
in an effective `luminosity factor',
$L(x_1,\beta,x_{\P};k,M_{jj})$.
That is the final cross section reads
\begin{equation}
\label{L-sig} \frac{M^2_{jj}d\sigma}{dY_{jj}dM^2_{jj}d\beta dk^2}=
L(x_1,\beta,x_{\P};k,M_{jj})\frac{d\hat\sigma}{d\hat t}\ ,
\end{equation}
where $Y_{jj}$ is the rapidity of the high $E_T$ dijet system and
$x_{\P}$ is the fraction of the proton's momentum carried by the soft (upper) Pomeron.

Using the notation $I^{(s)}$ of eqs.~({\ref{20}-\ref{2q}) we can write the effective
luminosity as
\begin{equation}
\label{Ls}
 L^{(s)}(x_1,\beta,x_{\P};k,M_{jj})~=~\frac{\pi\alpha_s(k^2)}{4}I^{(s)}(k)
 f^a(x_{\P},\beta,k,\mu)S^2\ .
\end{equation}
Here  $f^{a}$ ($a=g,q$) denotes the unintegrated diffractive parton
distribution produced by the soft (upper) Pomeron; and
 $\beta$ is the fraction of the soft Pomeron's momentum carried by the parton $k$.
 The factor $S^2$ is the soft gap survival factor which accounts for the absorptive
 corrections (see, for example, the review in \cite{Sgap}). In other words $S^2$
 is the probability that the LRG will not be filled by secondaries produced
 by additional inelastic interactions which may accompany the main process of Fig.3.

Note that by using integral (\ref{1}) in (\ref{Ls}) we have assumed that
the hard (lower) Pomeron is described just by the two-gluon
exchange diagram. To be more precise the two-gluon exchange factor
$1/((q^2_1+m^2_g)(q^2_2+m^2_g))$ should be replaced by the
unintegrated GPD function. That is, when calculating the $I^{(s)}(k)$
of eqs.~({\ref{20}-\ref{2q}) we have to use
\begin{equation}
\label{I-GPD} I_{\mu\nu}=\int d^2q\frac{q_{1\mu}\cdot
F_g(x_1,x_2;q,\mu)}{(q^2_1+m^2_g)(q^2_2+m^2_g)}
\left[\frac{k_{t,\nu}}{k^2+m^2_g}
-\frac{(k-q_2)_{t,\nu}}{(k-q_2)^2+m^2_g}\right]\
.
\end{equation}
The GPD function $F_g$ in our $x_2\ll x_1$ limit can be written in the
simplified form~\cite{Pros,KMR-pr}
\begin{equation}
\label{Fg} F_g(x_1,x_2;q,\mu)=R_g\frac{\partial}{\partial\ln
q^2}\left[\sqrt{T_g(q,\mu)}x_1g(x_1,q^2)\right]\ ,
\end{equation}
where $g(x_1,q^2)$ is the usual integrated gluon distribution and
$R_g$ is given by (\ref{Rg}). Here and below we put for simplicity
$Q=0$ and omit this argument from now on. Note that working in terms of
unintegrated distributions we have no explicit $T$-factor in
(\ref{I-GPD}). It is already included in the expressions for $f^a$
and $F_g$ (see (\ref{Fg})).

At first sight it looks as if the integral (\ref{I-GPD}) over $q^2$
has a logarithmic $dq^2/q^2$ form for $q^2\ll k^2$, and instead of the
unintegrated distribution (\ref{Fg}) one can use the
full GPD function taken at a scale equal to $k^2$.
However, this is not true. Expanding the expression in the square
brackets in (\ref{I-GPD}) over the $q/k$ ratio and averaging over
the azimuthal angle, we see that the first term proportional to $q/k$
vanishes, while the remaining terms do not have a logarithmic
structure. Simultaneously the integral over the lowest jet momentum
$k$ also has a non-logarithmic form.

 Note also that in the denominators $1/k^2$ and $1/(k-q_2)^2$, corresponding
 to a parton radiated from the soft Pomeron  in (\ref{I-GPD}), we have
 to keep the full parton virtuality
 $k^2=k^2_t/(1-\beta)$. Therefore, with the previous notation $k^2=k^2_t$
 we must replace (\ref{I-GPD}) by
\begin{equation}
\label{I-GPD-f} I_{\mu\nu}=\int d^2q\frac{q_{1\mu}\cdot
F_g(x_1,x_2;q,\mu)}{(q^2_1+m^2_g)(q^2_2+m^2_g)}
\left[\frac{(1-\beta)k_\nu}{k^2+q^2_{\P}}
-\frac{(1-\beta)(k-q_2)_\nu}{(k-q_2)^2+q^2_{\P}}\right]\
.
\end{equation}

Finally for the unintegrated soft Pomeron distribution $f^a$ we have
taken (\ref{unint-k}) from the fit B parametrization of the H1
collaboration~\cite{H1} assuming that at $Q^2_k<Q_0^2$ the values of
$a^D(x_{\P},\beta,Q_k;t)$ and $\partial
a^D(x_{\P},\beta,Q_k;t)/\partial \ln Q^2_k$ are frozen; that is, equal
to their value at $Q=Q_0$. For very small $Q_k<1$ GeV we put
$a^D(...)\propto Q^2_k$ but this negligibly changes the results in
comparison with the simple `frozen' assumption (recall that here
$Q^2_k=k^2_t/(1-\beta)$).
Note that in (\ref{I-GPD-f}) an effective infrared cutoff $q_{\P}$,
corresponding to the soft Pomeron size, is included in order to have  the
possibility of considering Pomerons with a size smaller than that
given by the cutoff $m_g$; we put $q_{\P}=1$ GeV  (or 0.2 GeV).


\section{Hard dijet $(2\to 2)$ cross section}
The cross section of hard subprocess of dijet production is
calculated using the MHV formalism~\cite{MP}. The only non-trivial
point is that now we are not looking for the usual colour-averaged
cross sections, but for cross sections with the high $E_T$ dijet in
either a colour-octet state (if the parton $k$ is a gluon) or a
colour-triplet state (if it is a quark). That is, the hard
matrix element $\cal{M}$ for the differential  cross section
\begin{equation}
\frac{d\hat\sigma}{dt}~=~\frac{|{\cal M}|^2}{16\pi\hat s^2}
\end{equation}
is given as follows:
\begin{equation}
{\cal M}^{acd}_{\lambda_c\lambda_d}(gg\to gg;\ J_z=0)=\frac
1{4(N^2_c-1)}\sum_{b,e}
f^{abe}\sum_{\lambda_e,\lambda_b}\delta_{\lambda_b\lambda_e}
M^{ebcd}_{\lambda_e\lambda_b\lambda_c\lambda_d}\
,
\end{equation}
\begin{equation}
{\cal M}^{acd}_{\lambda_c\lambda_d}(gg\to gg;\ J_z=2)=\frac
1{4(N^2_c-1)}\sum_{b,e}
f^{abe}\sum_{\lambda_e,\lambda_b}\delta_{\lambda_b-\lambda_e}
M^{ebcd}_{\lambda_e\lambda_b\lambda_c\lambda_d}\
,
\end{equation}
\begin{equation}
{\cal M}^{aik}_{\lambda_c\lambda_d}(gg\to q\bar q;\ J_z=0)=\frac
1{4(N^2_c-1)}\sum_{b,e}
f^{abe}\sum_{\lambda_e,\lambda_b}\delta_{\lambda_b\lambda_e}
M^{ebik}_{\lambda_e\lambda_b\lambda_i\lambda_k}\
,
\end{equation}
\begin{equation}
{\cal M}^{aik}_{\lambda_c\lambda_d}(gg\to q\bar q;\ J_z=2)=\frac
1{4(N^2_c-1)}\sum_{b,e}
f^{abe}\sum_{\lambda_e,\lambda_b}\delta_{\lambda_b-\lambda_e}
M^{ebik}_{\lambda_e\lambda_b\lambda_i\lambda_k}\
,
\end{equation}
and
\begin{equation}
{\cal M}^{ick}_{\lambda_i\lambda_b\lambda_c\lambda_k}(gq\to
gq)=\frac 1{2(N^2_c-1)}\sum_{b,i'}
t^b_{ii'}\sum_{\lambda_{i'}}\delta_{\lambda_i\lambda_{i'}}
M^{i'bck}_{\lambda_{i'}\lambda_b\lambda_c\lambda_k}\
.
\end{equation}
where $a,b,c,d,e=1,2,...,8$ ($i,i',k=1,2,3$) are the gluon (quark)
colour indices, while the $\lambda_a,\lambda_b,...=\pm 1$ are the
helicities of the corresponding gluon or quark ($\lambda_i=\pm
1/2$). $M^{ebcd}_{\lambda_e\lambda_b\lambda_c\lambda_d}(gg\to gg)$
and $ M^{ick}_{\lambda_{i'}\lambda_b\lambda_c\lambda_k}(qg\to qg)$
are the conventional matrix elements.
These formulae, with the unusual clour strcture exposed, are needed for the calculation of three (or more) jet production.
 
With this unusual colour structure we now find that the hard  cross sections, averaged over the colours and helicities
 of incoming partons and summed for the outgoing partons,  read:
\begin{eqnarray}
\frac{d\hat\sigma^{(0)}(gg\to gg)}{dt}&=&\frac{\pi\alpha^2_sN^3_c}{(N^2_c-1)^2}
\frac 1{p^4_T}\left(1-\frac{4p^2_T}{\hat s}\right)\ ,\\
\frac{d\hat\sigma^{(0)}(gg\to q\bar q)}{dt}&=& 0\ ,\\
\frac{d\hat\sigma^{(2)}(gg\to gg)}{dt}&=&\frac{\pi\alpha^2_sN^3_c}{(N^2_c-1)^2}
\frac 1{p^4_T}\left(1-\frac{4p^2_T}{\hat s}\right)(1-4p^2_T/\hat s+4p^4_T/\hat s^2),\\
\frac{d\hat\sigma^{(2)}(gg\to q\bar q)}{dt}&=&\frac{\pi\alpha^2_sN^2_c}{4(N^2_c-1)^2}
\frac 1{p^2_T\hat s}\left(1-\frac{4p^2_T}{\hat s}\right)(1-2p^2_T/\hat s)\ ,\\
\frac{d\hat\sigma^{q}(gq\to
gq)}{dt}&=&\frac{\pi\alpha^2_s}{N^3_c(N^2_c-1)}\left(\frac{2N^2_c\hat
u +\hat t}{\hat t}\right)^2\frac{\hat u^2+\hat s^2}{16\hat s\hat u}\
.
\end{eqnarray}
where 
\be
\hat s=M^2_{jj}, ~~~~~ \hat t=\hat s(1-\cos(\theta))/2,~~~~~
\hat u=\hat s(1+\cos(\theta))/2,
\ee
 are the Mandelstam variables
corresponding to the hard subprocess; $\theta$ is the scattering
angle in dijet rest system; and  $p^2_T=\hat t\hat u/\hat s$.

Note that in the case of gluon-gluon collisions the factor
$(1-4p^2_T/\hat s)$ vanishes at $\theta=\pi/2$.  This reflects the
fact that we deal with a $gg$ system in the asymmetric ($f^{abc}$ tensor)
colour-octet state. Therefore the corresponding (symmetric) $gg$
wave function  has a zero at $90^o$.

\section{Numerical example}
The above formulation allows the evaluation of the role of `soft-hard' Pomeron fusion as a background to CEP high $E_T$ jet production. As a numerical example we
calculate the cross section of central semi-exclusive dijet
production at $\sqrt s=13$~TeV for jets with $p_T=30$~GeV and rapidity of
dijet system $Y_{jj}=0$. We take the dijet scattering angle
$\theta=45^o$ (in dijet c.m.s.) in order not to affect the result by
the vanishing of gluon-gluon induced colour-octet cross sections
(21,23,24) at $\theta=90^o$. That is the two high $E_T$ jets are separated
by the pseudorapidity interval $\Delta\eta=3.5$ (corresponding to jets with $\eta_j=\pm 1.75$).
We sum over all types of parton jets. That is, a jet may be a gluon or a
light quark $(u,d,s,c)$ jet.\footnote{We take the survival factor $S^2=0.02$ as in model-2 of~\cite{KMR-13}.} 
Next 
 the dijet system is accompanied by a softer third jet,
allowing transverse momentum $k_3< p_T$. Specifically we consider $k_3<3,~6$ and 10~GeV.
In addition we allow radiation from the soft Pomeron with $k_i<k_3$.
The results are shown in Fig.4; we plot the distribution over the ratio
$M_{jj}/M_X=R_{jj}=\beta$ and compare this with the cross section
of pure exclusive dijet production (shown by the dashed line in the upper right corner).
For the `hard' Pomeron we use the integrated parton distributions of \cite{MMHT14} and for the `soft'
Pomeron) the diffractive parton distributions of H1 fit B~\cite{H1}~\footnote{The NLO DGLAP evolution of fit B input distributions
was performed by QCDNUM program~\cite{QCDNUM}.}.

\begin{figure} [h]
\vspace{-5.cm} 
\hspace{1.9cm}
\includegraphics[scale=0.6]{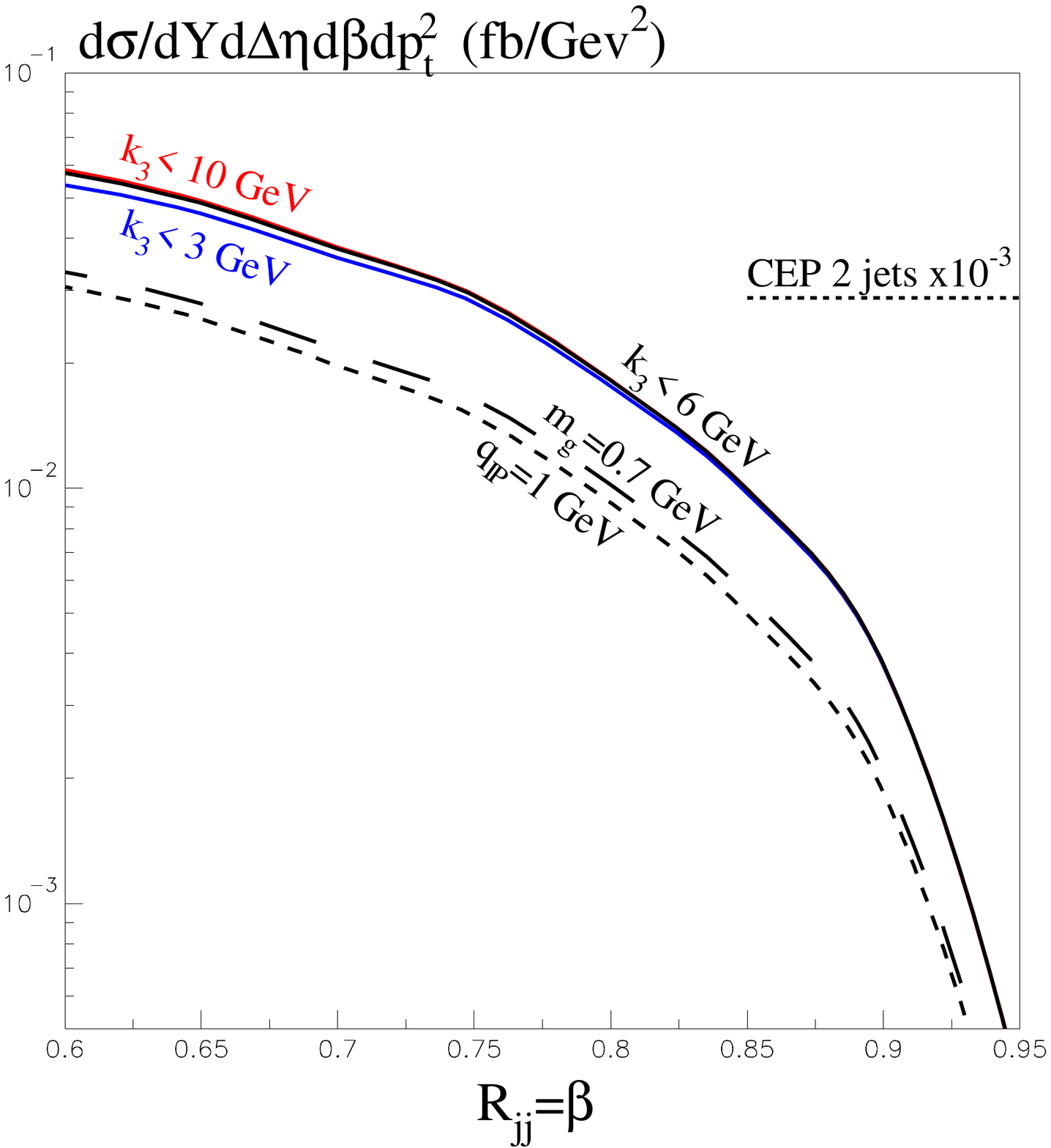}
\vspace{-.0cm} 
\caption{\sf The cross section of three jet
semi-exclusive central production integrated over the third jet
transverse momentum $k_3$ up to 3, 6 and 10 GeV (respectively shown by blue, black and red
continuous curves) at $\sqrt s=13$ TeV as the
function of the ratio $R_{jj}=M_{jj}/M_X$. The two high $E_T$ jets have
$p_T=30$ GeV and pseudorapidities $\eta_j=\pm 1.75$, $Y_{jj}=0$.
An infrared cutoff $m_g=q_{\P}=0.2$ GeV was used for the main
calculation shown by continuous curves. Two alternative choices of cutoff are also shown: $m_g=0.7$ GeV corresponding to
the long dashed curve with $k_3<6$ GeV, while $q_{\P}=1$ GeV and
$m_g=0.2$ GeV corresponds to the lower dashed curve ($k_3<6$ GeV).
The pure exclusive dijet cross section reduced by factor of 1000 is
shown by the horizontal short dashed line.  We used the integrated MMHT2014
parton distributions~\cite{MMHT14} (for the 'hard' Pomeron) and the
H1 fit B for the diffractive parton distributions~\cite{H1} for the soft Pomeron.}
\end{figure}

As emphasized above, the major contribution comes from the
relatively low transverse momenta of the third jet; the difference
between the blue ($k_3<3$ GeV) and the red ($k_3<10$ GeV) curves is
rather small. This fact justifies the possibility of calculating such
a 3-jet cross section using the factorization approach.

All the continuous curves were calculated using a weak infrared cutoff
of about 1 fm ($m_g=q_{\P}=0.2$ GeV). A stronger IR cutoff, shown by
the dashed curves,  reduces the cross section by a factor of about two.

In comparison with the pure exclusive (CEP) dijet production (shown by
the horizontal dotted line) the contribution of production driven by the soft-hard Pomeron fusion   mechanism is practically
negligible. It is smaller by three orders of magnitude.
Recall that the probability of radiation of a third jet from the hard
matrix element in hard-hard Pomeron (CEP) fusion is about
10\%~\cite{HLKR-3j}. The role of DPE dijet production was studied in~\cite{Lonbl}. There the authors  accounted for the fact that due to detector resolution and the jet searching algorithm, the CEP peak is washed out and has a maximum at $R_{jj}\simeq 0.9$.  Allowing for the initial-state shower, that is including the DPE contribution, the cross section obtained in~\cite{Lonbl} already at $R_{jj}=0.9$ exceeds the CEP component by more than a factor of 2.

Such a strong suppression  of `hard-soft' Pomeron fusion is caused by the asymmetric structure of the amplitude. Besides the $\alpha_s$ suppression (in which the small value of coupling is not compensated by the logarithmic transverse momentum integration $\int dk^2_t/k^2_t$) our hard matrix element has an additional smallness due to the fact that we are looking for the production of large $E_T$ jets in an asymmetric colour octet state. Recall that the elementary  dijet cross sections (21-24) vanish at $\theta=90^o$.  Thus the absence of large logarithms in the $k_t$ and $q_t$ integrals over the incoming parton momenta and the numerically small factors coming from the asymmetric angular integrations, result in a very small contribution of this mechanism to the final high $E_T$ jets cross section.

\section{Conclusion}

We have considered the possibility of semi-exclusive high $E_T$ dijet
production accompanied by a third jet with smaller transverse
momentum plus the possibility of additional radiation coming from 
soft Pomeron spectators. That is jet production from soft-hard Pomeron fusion.

 We have shown that the cross section of such a process can be calculated using
 the factorization of hard dijet cross sections and an effective
 luminosity which describes the probability to find appropriate incoming
 partons and to emit the third jet. Moreover the role of the infrared cutoff was studied.

 We found that the contribution of this channel is quite small in comparison
 with pure CEP dijet production. This fact, which was not evident a priori, greatly simplifies the calculation
 and interpretation of the exclusive  (and semi-exclusive) high $E_T$ jet production since one can neglect the hard-soft Pomeron fusion contribution.

Note that the cross section that we have calculated is just the simplest example of processes which may arise from  the fusion of  two different structures of the Pomeron. The result that processes caused by soft-hard Pomeron fusion can be factorized, as the convolution of a pure hard matrix element and an effective luminosity calculated at a much lower scale, has a universal nature.  That is, the effective luminosity can be applied to other central diffractive processes.  However, as we have shown in our 3-jet example, the cross section arising from hard-soft Pomeron fusion turns out to be small. Of course, in cases where the original CEP amplitude is suppressed, for example by the $J_z=0$ selection rule as in $b\bar{b}$ production, then the hard-soft fusion contribution may be noticeable.

\section*{Acknowledgements}
 
 MGR thanks the IPPP at the University of Durham for hospitality.
VAK acknowledges support from a Royal Society of Edinburgh Auber award.


\thebibliography{}
 \bibitem{St-Cand} 
L.A. Harland-Lang, V.A. Khoze, M.G. Ryskin, W.J. Stirling,
Eur.Phys.J. {\bf C69} (2010) 179-199.
 \bibitem{J0} 
Valery A. Khoze, Alan D. Martin, M.G. Ryskin,  Eur.Phys.J.{\bf C19}
(2001) 477-483, Erratum: Eur.Phys.J. {\bf C20} (2001) 599;\\
  V.~A.~Khoze, A.~D.~Martin and M.~G.~Ryskin,
  hep-ph/0006005 (2000).
   \bibitem{dur-rev} 
L.A. Harland-Lang, V.A. Khoze, M.G. Ryskin, W.J. Stirling,
Int.J.Mod.Phys.{\bf  A29} (2014) 1430031.
\bibitem{Pros}V.A. Khoze, A.D. Martin, M.G. Ryskin, Eur. Phys. J. {\bf C23} (2002) 311.
  \bibitem{Col} 
Arjun Berera, John C. Collins,  Nucl.Phys. {\bf B474} (1996) 183.
 \bibitem{KMR-dijet} V.A. Khoze, A.D. Martin, M.G. Ryskin, Phys. Rev. {\bf D56} (1997) 5867.

 \bibitem{dur-rev1} 
M.G. Albrow, T.D. Coughlin, J.R. Forshaw, Prog.Part.Nucl.Phys. {\bf 65} (2010) 149. 
 \bibitem{CDF}
CDF Collaboration (T. Aaltonen et al.), Phys.Rev. {\bf D77} (2008)
052004.
\bibitem{DPE} V.A. Khoze, A.D. Martin, M.G. Ryskin, Eur. Phys. J. {\bf C48} (2006) 467.
 \bibitem{H1}
H1 Collaboration (A. Aktas  et al.) Eur.Phys.J.{\bf  C48} (2006) 715-748,\\
H1 Collaboration (A. Aktas et al.) Eur.Phys.J. {\bf C48} (2006)
749-766.
 \bibitem{Zeus} 
ZEUS Collaboration (S. Chekanov et al.). Eur.Phys.J. {\bf C38}
(2004) 43-67.
 \bibitem{PR} 
M. Boonekamp, Robert B. Peschanski, C. Royon,  Phys.Rev.Lett. {\bf
87} (2001) 251806.
\bibitem{POMW}  B.E. Cox and J.R. Forshaw, Compute. Phys. Commun. {\bf 144} (2002) 104.
\bibitem{HLKR-3j}
L.A. Harland-Lang, V.A. Khoze, M.G. Ryskin, Eur.Phys.J. {\bf C76}
(2016) 9 (sect.4.1).
\bibitem{Lonbl}  	
Leif Lonnblad, Radek Zlebcík, Eur.Phys.J. {\bf C76} (2016) 668.
\bibitem{PP}
G. Parisi, R. Petronzio, Phys.Lett. 94B (1980) 51-53 ($m_g=0.8$
GeV).

\bibitem{CF}
M. Consoli, J.H. Field, Phys.Rev. D49 (1994) 1293-1301.
($m_g=0.66\pm 0.08$ GeV).
\bibitem{KMR1} V.A. Khoze, A.D. Martin, M.G. Ryskin,  Phys. Lett. {\bf B401} (1997) 330.
\bibitem{MP} Michelangelo L. Mangano, Stephen J. Parke,  Phys.Rept. {\bf 200} (1991) 301.


\bibitem{MRW} A.D. Martin, M.G. Ryskin, G. Watt, Eur. Phys. J. {\bf C66} (2010) 163.
\bibitem{TF}
T.D. Coughlin, J.R. Forshaw,  JHEP {\bf 1001} (2010) 121.
\bibitem{updf} 
M.A. Kimber, Alan D. Martin, M.G. Ryskin, Phys.Rev. {\bf D63} (2001)
114027.
\bibitem{LN} F.E. Low, Phys. Rev. {\bf D12} (1975) 163;\\
S. Nussinov, Phys. Rev. Lett. {\bf 34} (1975) 1280.
\bibitem{GPD}   M.~Diehl,
  Phys.\ Rept.\  {\bf 388} (2003) 41. 

\bibitem{Shuv}
A.G. Shuvaev, Krzysztof J. Golec-Biernat, Alan D. Martin, M.G. Ryskin, Phys.Rev. {\bf D60} (1999) 014015;\\
A. Shuvaev, Phys.Rev. {\bf D60} (1999) 116005.
\bibitem{Sgap}  V.A. Khoze, A.D. Martin, M.G. Ryskin, J.Phys. {\bf G45} (2018)  053002.
\bibitem{KMR-pr}  V.A. Khoze, A.D. Martin, M.G. Ryskin, Eur. Phys. J. {\bf C14} (2000) 525.
 \bibitem{KMR-13}  V.A. Khoze, A.D. Martin, M.G. Ryskin, Eur. Phys. J. {\bf C73} (2013) 2503.
 \bibitem{MMHT14} 
L.A. Harland-Lang, A.D. Martin, P. Motylinski, R.S. Thorne,
Eur.Phys.J. {\bf C75} (2015) 204.
\bibitem{QCDNUM} 
Botje, Comput. Phys. Commun. {\bf 182} (2011) 490, arXiv:1005.1481,
Erratum arXiv:1602.08383 (2016).
 \end{document}